\documentstyle[11pt]{article}

\begin{document}
\medskip                         
\centerline{{\Large\bf The dispersive self-dual Einstein equations}}

\vspace{.2in}\hfill\break
\centerline{{\Large\bf and the Toda Lattice}}

\vspace{.2in}\hfill\break
\centerline{{\bf I.A.B. Strachan}}

\vspace{.1in}\hfill\break
\centerline{Dept. of Pure Mathematics and Statistics, University of Hull,}
\vspace{.1in}\hfill\break
\centerline{Hull, HU6 7RX, England.}
\vspace{.1in}\hfill\break
\centerline{e-mail: i.a.b.strachan@hull.ac.uk}

\vspace{.3in}\hfill\break
\centerline{{\bf Abstract}}

\vspace{.2in}
\small
\parbox{5.8in}{The Boyer-Finley equation, or $SU(\infty)$-Toda equation
is both 
a reduction of the self-dual Einstein equations and the dispersionless
limit of 
the $2d$-Toda lattice equation. This suggests that there should be a
dispersive 
version of the self-dual Einstein equation 
which both contains the Toda lattice equation and whose dispersionless 
limit is the familiar self-dual Einstein equation. Such a 
system is studied in this paper. The results are achieved by using a 
deformation, based on an associative $\star$-product, of the algebra 
$sdiff(\Sigma^2)$ used in the study of the undeformed, or dispersionless, 
equations.}

\normalsize
\bigskip
\section*{1. Introduction }

One much studied $(2+1)$-dimensional integrable system, known as the
Boyer-Finley 
equation \cite{FinleyPlebanski,BoyerFinley}, is

\begin{equation}
\nabla^2 \rho=\partial_x^2  e^{\rho} \,,
\label{eq:BoyerFinley}
\end{equation}

\noindent where $\rho=\rho(z,\tilde{z},x)$ and 
$\nabla^2=\partial_z\partial_{\tilde{z}}\,.$
This also has been referred to, for reasons that will be explained 
latter, as the $SU(\infty)$-Toda or $sdiff(\Sigma^2)$-Toda equation. This 
equation has two interesting properties. Firstly it is a reduction of the 
self-dual Einstein equations (the equations governing any metric with
self-dual 
Weyl tensor and vanishing Ricci tensor) under a rotational
Killing vector. Secondly it is the dispersionless (or long wave or
continuum) 
limit of the Toda lattice equation

\begin{equation}
\nabla^2 \rho_n = e^{\rho_{n+1}} - 2 e^{\rho_n} + e^{\rho_{n-1}} \,,
\label{eq:TodaLattice}
\end{equation}

\noindent where $\rho_n=\rho_n(z,\tilde{z})\,.$ 
Heuristically, solutions whose natural length scale is large compared 
with the lattice spacing do not see the lattice, and the difference
operator 
on the right hand side may be approximated by a second derivative.
These two 
properties may be summarised in the following diagram:

\bigskip

$$\def\normalbaselines{\baselineskip20pt
\lineskip3pt \lineskiplimit3pt }
\def\mapright#1{\smash{
\mathop{\longrightarrow}\limits^{#1}}}
\def\mapdown#1{\Big\downarrow
\rlap{$\vcenter{\hbox{$\scriptstyle#1$}}$}}
\matrix{
&& \pmatrix{{\rm Toda~Lattice}\cr {\rm equation} \cr
{\rm [3-dimensions]}\cr}\cr
&&\cr
&&\mapdown{}& \hidewidth{\rm ~~dispersionless~limit} \hidewidth\cr
&&\cr
\pmatrix{{\rm self-dual~Einstein}\cr
{\rm equation}\cr{\rm [4-dimensions]}\cr} & \mapright{} & 
\pmatrix{{\rm Boyer-Finley}\cr {\rm equation}\cr
{\rm [3-dimensions]}\cr} & \cr
&\hidewidth{\rm reduction}\hidewidth& & \cr}
$$

\bigskip        

\noindent This diagram suggests that there should be some system whose
dispersionless 
limit is the self-dual Einstein equation, and which contains the Toda
lattice 
equation embedded within it as a special case. This system, the subject
of this 
paper, will be called the dispersive self-dual Einstein equation
(such a system
may not be unique, see section 5). This will 
have exactly those properties needed to make the following diagram
commute:

\bigskip

$$\def\normalbaselines{\baselineskip20pt
\lineskip3pt \lineskiplimit3pt }
\def\mapright#1{\smash{
\mathop{\longrightarrow}\limits^{#1}}}
\def\mapdown#1{\Big\downarrow
\rlap{$\vcenter{\hbox{$\scriptstyle#1$}}$}}
\matrix{
&&\hidewidth{\rm reduction}\hidewidth& & \cr
&\pmatrix{{\rm dispersive~self-dual}\cr{\rm Einstein~equation}\cr 
{\rm [4-dimensions]}\cr}&\mapright{}& \pmatrix{{\rm Toda~Lattice}\cr 
{\rm equation} \cr{\rm [3-dimensions]}\cr}\cr
&&&\cr
\hidewidth{\rm dispersionless~limit}\hidewidth&\mapdown{}&&\mapdown{}&
\hidewidth{\rm ~~dispersionless~limit} \hidewidth\cr
&&&\cr
&\pmatrix{{\rm self-dual~Einstein}\cr {\rm equation}\cr
{\rm [4-dimensions]}\cr} & \mapright{} & 
\pmatrix{{\rm Boyer-Finley}\cr {\rm equation}\cr
{\rm [3-dimensions]}\cr} & \cr
&&\hidewidth{\rm reduction}\hidewidth& & \cr}
$$

\bigskip

\centerline{\hbox{\bf Figure 1}}

\bigskip

\noindent Such a system was introduced by the author \cite{StrachanA} 
and studied further in \cite{Takasaki}, \cite{Castro}
and \cite{Przanowski}. These papers, 
however, do not study the relationship between the 
dispersive self-dual Einstein equation and the Toda lattice equation.

\section*{2. The self-dual Einstein equations}

There are several ways to define the equations governing a self-dual
vacuum 
metric. One particular useful method is the following \cite{MasonNewman}. 
Let ${\cal U}=V_1+\lambda V_2$ and ${\cal V}=V_3 + \lambda V_4$ be two 
commuting vectors fields (for all values of the spectral
parameter $\lambda\,,$
and where the $V_i$ are independent of $\lambda$) on some four-manifold 
$\cal M\,.$  Suppose further that each $V_i$ preserves a
four-form $\omega$ on 
$\cal M\,.$ Then the contravariant metric

\begin{equation}
{\bf g} = \Lambda^{-1} [ V_1 \otimes_S V_4 - V_2 \otimes_S V_3 ] \,,
\label{eq:metric}
\end{equation}

\noindent where the conformal factor is defined by

\[
\Lambda=\omega(V_1,V_2,V_3,V_4)\,,
\]

\noindent is a self-dual metric. It may also be shown that all self-dual 
metrics may be written in this way \cite{MasonNewman}.
One approach that has attracted much 
attention is to write ${\cal M} = R^2 \times \Sigma^2$ and to take
the vectors 
fields to belong to $sdiff(\Sigma^2)$, the Lie algebra of volume
preserving 
diffeomorphisms of the 2-surface $\Sigma^2\,.$ Explicitly, write

\[
X_f = {\partial f(z,\tilde{z},x,p)\over \partial x} 
{\partial\phantom{p}\over\partial p} - 
{\partial f(z,\tilde{z},x,p)\over \partial p} 
{\partial\phantom{x}\over\partial x} \,.
\]

\noindent Here $z$ and $\tilde{z}$ are coordinates on $R^2$ and $x$
and $p$ are 
coordinates on $\Sigma^2\,.$ These vector fields obey the important
relation
${[} X_f,X_g {]} = X_{\{f,g\}}\,,$
where $\{f,g\}$ is the Poisson bracket

\[
\{f,g\} = \frac{\partial f}{\partial x}\frac{\partial g}{\partial p}-
\frac{\partial f}{\partial p}\frac{\partial g}{\partial x}\,.
\]

With the vector fields

\begin{eqnarray*}
{\cal U} & = & \partial_z + \lambda X_f \,, \\
{\cal V} & = & \partial_{\tilde{z}} + \lambda X_g  
\end{eqnarray*}

\noindent and four-form
$\omega=dz \wedge d{\tilde{z}} \wedge dx \wedge dp$
one obtains Plebanski's form of the self-duality
equations \cite{Plebanski}:

\[
\Omega_{zx} \Omega_{{\tilde z}p} - \Omega_{zp} \Omega_{{\tilde z}x} = 1\,,
\]

\noindent or, using the Poisson bracket,

\begin{equation}
{ \{ \Omega_z , \Omega_{\tilde z} \} } = 1 \,.
\label{eq:PlebanskiOne}
\end{equation}

\smallskip

\noindent Other choices lead to other, equivalent, forms of
the self-duality equations.
For example, with the vector fields (where the functions
$f\,,g$ and $h_{\pm}$
depend on the four variables
$z\,,\tilde z\,,x$ and $p$)

\begin{eqnarray*}
V_1 & = & \phantom{-} \partial_z + X_f \,,\\
V_2 & = & \phantom{-} X_{h_{+}}\,,\\
V_3 & = & \phantom{-} X_{h_{-}}\,,\\
V_4 & = & - \partial_{\tilde z} - X_g
\end{eqnarray*}

\noindent which preserve the four-form

\[
\omega= d \tilde z \wedge dz \wedge dx \wedge dp
\]

\noindent one obtains, from the condition the vector fields
${\cal U}=V_1+\lambda V_2$ and ${\cal V}=V_3+\lambda V_4$ commute for
all values of $\lambda\,,$ the equations

\begin{eqnarray}
\partial_z h_{-} + \{ f,h_{-} \} & = & 0 \,, \label{eq:sdA} \\
\partial_{\tilde z} h_{+} + \{ g , h_{+} \} & = & 0 \,, \label{eq:sdB} \\
\partial_z g - \partial_{\tilde z} f +\{f,g\} - \{ h_{+} , h_{-} \}
& = & 0 \,.
\label{eq:sdC}
\end{eqnarray}

\noindent Hence, using the theorem proved in \cite{MasonNewman}
(see also \cite{WardA}),
any solution of this system generates a self-dual metric, one
particular form of which
is given by (\ref{eq:metric}).
There is obviously much freedom in this system,
there being three equations for the four functions, but this may be
removed by
fixing the gauge freedom (see \cite{MasonNewman}). This is entirely
analogous to the more familiar
self-dual Yang-Mills equations
$F_{\mu\nu}=\frac{1}{2}\epsilon_{\mu\nu\alpha\beta} F^{\alpha\beta}$,
which are also a set of
three equations for the four gauge potentials. Once again this freedom
may be removed by
fixing the gauge.

\bigskip

The Boyer-Finley equation (\ref{eq:BoyerFinley}) was first obtained by
studying the reductions of the 
self-duality equations under a non-self-dual Killing vector (if the
Killing vector 
is self-dual then the equations reduce to the three-dimensional Laplace 
equation) \cite{BoyerFinley}.
This result was obtained by performing a dimensional reduction on
Plebanski's
equation (\ref{eq:PlebanskiOne}) followed by a Legendre transformation.

\bigskip

One possible Lax pair for this equations, which clearly shows 
that this equation is a reduction of the self-dual Einstein equations, was
proposed by Ward \cite{Ward}. One may use the same vector fields

\begin{eqnarray*}
{\cal U} & = & (\partial_z + X_f) + \lambda X_{h_{+}} \,, \\
{\cal V} & = & X_{h_{-}} - \lambda (\partial_{\tilde{z}} + X_g) \,,
\end{eqnarray*}

\noindent as before, but with the functions satisfying the following
ansatz:

\begin{eqnarray*}
f & = & f(z,\tilde z,x) \,, \\
g & = & g(z,\tilde z,x) \,, \\
h_{\pm} & = & h(z,\tilde z,x) \exp (\pm p)\,,
\end{eqnarray*}

\noindent i.e. the functions $f$ and $g$ are independent of the
variable $p$ and
the functions $h_{\pm}$ have a specific dependence on this variable.
The condition that the two vector 
fields $\cal U$ and $\cal V$ commute implies the following set of
equations (note that with this ansatz the variable $p$ has dropped
out leaving equations in the remaining $z,\tilde z$ and $x$ variables):

\begin{eqnarray}
\partial_z h - h \partial_x f & = & 0 \,,\label{eq:BFa} \\
\partial_{\tilde{z}} h + h \partial_x g & = & 0 \,, \label{eq:BFb} \\
\partial_z g - \partial_{\tilde{z}} f+2 h \partial_x h & = & 0\,.
\label{eq:BFc} 
\end{eqnarray}

\noindent These reduce to the Boyer-Finley equation (\ref{eq:BoyerFinley}),
where $\rho=2\log h\,.$
Owing to the use of the algebra $sdiff(\Sigma^2)$ this equation
has also been 
called the $sdiff(\Sigma^2)$-Toda equation or the $SU(\infty)$-Toda
equation. This
establishes the bottom line in Figure 1.

\section*{3. Deformations of $\bf{sdiff({\Sigma^2})}$ }

Central to the above derivation is the identity 
${[} X_f,X_g {]} = X_{\{f,g\}}\,.$ The deformation that will be
studied here is 
to replace the Poisson bracket by the Moyal bracket \cite{Moyal}

\begin{eqnarray}
\{ f,g\}_M = \sum_{s=0}^\infty
{\hbar^{2s}\over 2^{2s} (2s+1)!} \sum_{j=0}^{2s+1} (-1)^j
\pmatrix{ 2s+1 \cr j}\,\,
(\partial_x^{2s+1-j}\partial_p^j f) \,\,
(\partial_x^j \partial_p^{2s+1-j} g)\,.
\label{eq:Moyal}
\end{eqnarray}

\noindent This has the important properties:

\begin{eqnarray*}
\{f,g\}_M & = & {f\star g-g\star f\over \hbar} \,, \\
\lim_{\hbar\rightarrow 0}   \{f,g \}_M & = & \{ f,g \} \,, \\
\{ \{ f,g \}_M , h \}_M + \{ \{g,h\}_M,f\}_M + \{ \{ h,f\}_M,g\}_M
& = & 0 \,,
\end{eqnarray*}

\noindent where $\star$ is an associative, though non-commutative, product 
which reduces to normal multiplication in the limit $\hbar\rightarrow 0$
defined by

\[
f \star g = \sum_{s=0}^\infty \frac{ \hbar^s}{2^s s!} \sum_{j=0}^s (-1)^j
\pmatrix{s \cr j} \,
(\partial_x^{s-j}\partial_p^j f) \,\, (\partial_x^j \partial_p^{s-j} g)\,.
\]

\noindent Such deformations are essentially unique \cite{unique};
any other deformation being 
ultimately equivalent to the Moyal bracket (\ref{eq:Moyal}). 
One such equivalent deformation is given by 

\begin{equation}
\{ f , g \}_K = \sum_{s=1}^\infty
{\hbar^{s-1} \over s!}
(\partial_x^s f \partial_p^s g-\partial_p^s f \partial_x^s 
g)\,,
\label{eq:Manin}
\end{equation}

\noindent this being the algebra of symbols of pseudo-differential
operators
(see, for example \cite{StrachanPreprint} and the reference therein)
and is known
as the Kuperschmidt-Manin bracket. This too may be written in terms of an
associative product, namely

\[
f \star g = \sum_{s=0}^\infty \frac{\hbar^s}{s!} \partial_x^s f
\,\,\partial_p^s g\,.
\]

\bigskip

The dispersive self-dual Einstein equation is defined by using such
brackets 
in place of the Poisson bracket in equation (\ref{eq:PlebanskiOne}) and/or
(\ref{eq:sdA})-(\ref{eq:sdC}). Such a procedure may be made more
geometrical 
in terms of a deformed differential geometry, but such an approach
would take 
one beyond the scope of this paper \cite{StrachanPreprint}.
As before, there are many different forms that this dispersive self-dual 
Einstein equation can take. One being \cite{StrachanA}

\begin{equation}
\{ \Omega_z, \Omega_{\tilde{z}} \}_M = 1\,,
\label{eq:Plebanski}
\end{equation}

\noindent this being a dispersive version of Plebanski's first heavenly 
equation. Further properties of this equation have been found 
\cite{Takasaki,Castro}. In particular, 
it has been shown that solutions may be encoded via a
Riemann-Hilbert problem 
in the corresponding Moyal loop group.
These results show the integrability of
such Moyal deformations.

\bigskip

The version used in this paper follows from
equations (\ref{eq:sdA})-(\ref{eq:sdC})
with the Poisson bracket replaced by the Moyal bracket:

\begin{eqnarray}
\partial_z h_{-} + \{ f , h_{-} \}_M & = & 0 \,, \label{eq:deformA} \\
\partial_{\tilde z} h_{+} + \{ g , h_{+} \}_M & = & 0 \,,
\label{eq:deformB} \\
\partial_z g - \partial_{\tilde z} f +
\{f,g\}_M - \{ h_{+} , h_{-} \}_M & = & 0 \,.
\label{eq:deformC}
\end{eqnarray}

\bigskip

\noindent These equations follow from the Lax pair
$[{\cal U} , {\cal V}] = 0 $
with $[ X_f , X_g ] = X_{ \{f,g\}_M}\,.$ The methods used to show
the integrability
of equation (\ref{eq:Plebanski}) may be used to prove the
integrability of this
four dimensional system.

This is the system which appears in the top left hand
corner in Figure 1. The property
$\lim_{\hbar\rightarrow 0} \{f,g\}_M = \{f,g\}\,$ of the Moyal bracket
establishes the left
hand side of Figure 1 (as in this limit
equations (\ref{eq:deformA}-\ref{eq:deformC}) reduce to equations
(\ref{eq:sdA}-\ref{eq:sdC})).
It remains to show that this system contains, via a
dimensional reduction, the Toda Lattice equation.

\section*{4. Reduction to the Toda Lattice equation}

In section 2 it was shown that with the ansatz

\begin{eqnarray*}
f & = & f(z,\tilde z,x) \,, \\
g & = & g(z,\tilde z,x) \,, \\
h_{\pm} & = & h(z,\tilde z,x) \exp(\pm p)
\end{eqnarray*}

\noindent the four dimensional self-dual Einstein
equations (\ref{eq:sdA})-(\ref{eq:sdC}) reduce to
the three dimensional Boyer-Finley equation. In this section the same
ansatz will be used in the
Moyal deformed version of these equations, the dispersive Einstein
equations
(\ref{eq:deformA})-(\ref{eq:deformC}), and, as before, the variable $p$
will drop out leaving equations in the remaining three variables.
At first sight this 
might seem a rather strange thing to do, as this will introduce
an infinite
number of extra 
terms into the governing equations, corresponding to the infinite
summation in
the definition of the Moyal bracket. However, it will turn out that it is 
possible to sum these infinite series and hence obtain results in
closed form.
The analogues of equations (\ref{eq:BFa}-\ref{eq:BFc}) are

\begin{eqnarray*}
\partial_z h-h\sum_{s=0}^\infty {\hbar^{2s}\over 2^{2s} (2s+1)!}
\partial_x^{2s+1} f & = & 0 \,,  \\
\partial_{\tilde{z}}h+h\sum_{s=0}^\infty {\hbar^{2s}\over 2^{2s} (2s+1)!}
\partial_x^{2s+1} g & = & 0 \,,  \\
\partial_z g - \partial_{\tilde{z}} f  
+\sum_{s=0}^\infty {\hbar^{2s}\over 2^{2s} (2s+1)!} \sum_{j=0}^{2s+1}
\pmatrix{ 2s+1 \cr j}
\partial_x^{2s+1-j} h \,\, \partial_x^j h  & = & 0
\end{eqnarray*}

\noindent and these may be resummed in terms of a difference operator

\begin{eqnarray*}
\partial_z h(x) - h(x) {f(x+\hbar/2) - f(x-\hbar/2) \over \hbar}
& = & 0 \,, \\
\partial_{\tilde{z}} h(x) + h(x) {g(x+\hbar/2) - g(x-\hbar/2) \over \hbar}
& = & 0 \,, \\
\partial_z g(x) - \partial_{\tilde{z}} f(x) +
{h^2(x+\hbar/2) - h^2(x-\hbar/2) \over\hbar} & = & 0
\end{eqnarray*}

\noindent (the dependence of these functions on the $z,\tilde{z}$
coordinates 
has been suspended for notational convenience).
With $\rho=2\log h$ one obtains the 
Toda Lattice equation

\begin{equation}
\nabla^2 \rho(x) =
\hbar^{-2} [e^{\rho(x+\hbar)} - 2 e^{\rho(x)} + e^{\rho(x-\hbar)} ]\,.
\label{eq:differenceTodaLattice}
\end{equation}

\noindent So the same ansatz used to show that the Boyer-Finley equation
is a reduction of the self-dual Einstein equations
may also be used to show
that the Toda Lattice is a reduction of the dispersive
self-dual Einstein equations. This establishes the top line in figure 1.
On
taking the
limit $\hbar\rightarrow 0$ one recovers the Boyer-Finley equation, which
is the remaining right hand side of figure 1.

As remarked earlier, the Boyer-Finley equation was obtained directly from
a dimensionally reduced version of Plebanski's equation
(\ref{eq:PlebanskiOne})
via a Legendre transformation. One might hope that there might be a direct
transformation from a dimensionally reduced version of
(\ref{eq:Plebanski}) to
the Toda Lattice 
(\ref{eq:differenceTodaLattice}), via some sort of Legendre transformation 
\cite{BoyerFinley,Castro}.

These equations differ in one important respect to the normal form of the 
Toda-Lattice equation (\ref{eq:TodaLattice});
the variable $x$ is continuous, not discrete.
As pointed 
out by Kupershmidt \cite{KupershmidtA}, taking the lattice to
be $\bf Z$ is not essential, and the 
main properties will of the Toda lattice equation will hold in
more general 
situations. 
For fixed $\hbar$ (say $\hbar=1$) one may embed any solution if the Toda 
lattice equation (\ref{eq:TodaLattice}) in the above equation 
(\ref{eq:differenceTodaLattice}), for example, for $x\in{\bf Z}$ let 
$\rho(z,{\tilde{z}},x)=\rho_x(z,{\tilde{z}})$ and $\rho=0$ elsewhere.
However, 
whether such solutions can be \lq smeared\rq~over the whole of the
domain of 
$x$, and how such solutions behave in the $\hbar\rightarrow 0$ limit
requires 
careful analysis. Also, in general one cannot obtain a specific
solution to
the Boyer-Finley equation from a specific solution of the Toda Lattice by
scaling and taking the $\hbar\rightarrow 0$ limit. However,
one particularly simple class of solutions for which this procedure
does work
is given by the ansatz

\[
\rho(z,\tilde{z},x) = \tilde\rho(z,\tilde{z}) + 2 \log x
\]

\noindent where $\tilde\rho$ satisfies Liouville's equation. This satifies
both the Boyer-Finley and the Toda lattice equations.

Using the alternative deformation of the Poisson bracket (\ref{eq:Manin}) 
one still obtains 
the Toda lattice equation, but by a slightly different route. The
analogues of 
equations (\ref{eq:BFa}-\ref{eq:BFc}) are

\begin{eqnarray*}
\partial_z h(x) - h(x) {f(x) - f(x-\hbar) \over \hbar} & = & 0 \,, \\
\partial_{\tilde{z}} h(x) + h(x) {g(x+\hbar)-g(x) \over \hbar}
& = & 0 \,, \\
\partial_z g(x) - \partial_{\tilde{z}} f(x)+ h(x) {h(x+\hbar) - h(x-\hbar) 
\over \hbar} & = & 0 \,,
\end{eqnarray*}

\noindent with which one obtains (\ref{eq:differenceTodaLattice}), with 
$\rho=\log h(x) + \log h(x+\hbar)$. Thus with the 
Moyal bracket one has a central difference operator
and with the Kupershmidt-Manin bracket (\ref{eq:Manin}) 
one has a forward/difference operator. Note that in both 
cases the deformation parameter $\hbar$ plays the role of the 
lattice spacing. This equivalence (between the forward/backward difference
scheme and the central difference scheme) is well know and just
corresponds to a gauge transformation. Here we see that this
equivalence may
also to traced back to the uniqueness theorems for
associative $\star$-products \cite{unique}.

\section*{5. Comments}

The Toda-lattice equation (\ref{eq:TodaLattice}) can be interpreted as the 
large $N$-limit of the $SU(N)$-Toda equation

\begin{equation}
\nabla^2 \rho_a = \sum_b K_{ab} e^{\rho_b}\,,
\label{eq:Toda}
\end{equation}

\noindent where $K_{ab}$ is the Cartan matrix of the $SU(N)\,.$
For finite $N$, these equations may be embedded within the self-dual 
Yang-Mills equations with gauge group $SU(N)\,.$ However, taking the large 
$N$-limit is problematical, especially when one tries to understand the 
integrability of these equations. For example,
the Riemann-Hilbert problems 
used to construct solutions to the Toda lattice equation is very
different in 
character to that used for the above, finite, Toda equation
(for a review, see 
\cite{AblowitzClarkson}). Alternatively, but 
presumably equivalently, the geometric constructions behind the above Toda 
equation and the Boyer-Finley equations are also somewhat different. The
results of this paper suggests one should study an interim equation, the 
difference-Toda equation (\ref{eq:differenceTodaLattice}), to understand 
the relationships between these equations; the limit that takes one 
from (\ref{eq:Toda}) to (\ref{eq:BoyerFinley}) being a double limit, first 
$N\rightarrow\infty$ then $\hbar\rightarrow 0\,.$ Similar results apply to 
understanding the relationship between the KP equation and the
dispersionless 
KP equation \cite{StrachanB,KupershmidtB}.

These results show the connection between the Moyal algebra and the
algebra
of difference operators normally used in the study of the infinite
Toda Lattice.
This relationship has been noticed recently, in various different
contexts, by
a number of authors and is the starting point for the construction of a
deformed differential calculus \cite{StrachanPreprint,misc}.

Finally, it must be noted that while all forms of the dispersionless
self-dual
Einstein equations are equivalent (i.e. one may go from any self-dual
metric to
any other via a coordinate transformation), it is by no means certain
that their
dispersive counterparts are so equivalent
(e.g. while (\ref{eq:PlebanskiOne})
and (\ref{eq:sdA}-\ref{eq:sdC}) are equivalent, it is not obvious
that (\ref{eq:Plebanski}) and (\ref{eq:deformA}-\ref{eq:deformC}) are
equivalent), and so the definite article in the title of this paper is,
perhaps,
premature. Thus what has been proved in this paper is the one
particular form of the
dispersive self-dual Einstein equations (one from from within a
large class of possible
forms) contains the Toda lattice as a dimensional reduction.
The equivalence, or
otherwise, of these dispersive self-dual Einstein equations
clearly deserves
closer attention.

\section*{Acknowledgements}

This paper was written while the author was the Wilfred Hall
Research Fellow
at the University of Newcastle. I would also like to thank the
referees for
their helpful comments.

\end{document}